# A COMPARATIVE STUDY OF THE FORMATION OF AROMATICS IN RICH METHANE FLAMES DOPED BY UNSATURATED COMPOUNDS

H.A. GUENICHE, J. BIET, P.A. GLAUDE, R. FOURNET, F. BATTIN-LECLERC[*]

*Département de Chimie-Physique des Réactions,*

*Nancy Université, CNRS,*

*1 rue Grandville, BP 20451, 54001 NANCY Cedex, France*

---

[*] E-mail : Frederique.battin-Leclerc@ensic.inpl-nancy.fr ; Tel.: 33 3 83 17 51 25 , Fax : 33 3 83 37 81 20

**ABSTRACT**

For a better modeling of the importance of the different channels leading to the first aromatic ring, we have compared the structures of laminar rich premixed methane flames doped with several unsaturated hydrocarbons: allene and propyne, because they are precursors of propargyl radicals which are well known as having an important role in forming benzene, 1,3-butadiene to put in evidence a possible production of benzene due to reactions of $C_4$ compounds, and, finally, cyclopentene which is a source of cyclopentadienylmethylene radicals which in turn are expected to easily isomerizes to give benzene. These flames have been stabilized on a burner at a pressure of 6.7 kPa (50 Torr) using argon as dilutant, for equivalence ratios ($\phi$) from 1.55 to 1.79. A unique mechanism, including the formation and decomposition of benzene and toluene, has been used to model the oxidation of allene, propyne, 1,3-butadiene and cyclopentene. The main reaction pathways of aromatics formation have been derived from reaction rate and sensitivity analyses and have been compared for the three types of additives. These combined analyses and comparisons can only been performed when a unique mechanism is available for all the studied additives.

# INTRODUCTION

While the formation of PAHs and soot has represented an important area of interest for kineticists for two decades, questions still remain concerning even the formation and the oxidation of the first aromatic compounds. In previous studies, the formation of benzene is mostly related to the reactions of $C_2$ (acetylene), $C_3$ or $C_4$ unsaturated species [1-7]. Nevertheless, some papers suggest a link between $C_6$ and $C_5$ cyclic species [8-11]. In a recent paper [12], we have investigated the reactions, in a rich methane flame, of allene and propyne because they are precursors of propargyl radicals which have an important role in forming benzene. In a second part [13], we have analyzed the reactions of 1,3-butadiene under similar conditions and put in evidence an important route to the production of benzene due to reactions of $C_4$ compounds. To finish this work, we have studied the reactions of cyclopentene which is a source of $C_5$ radicals [14]. The purpose of the present paper is to compare the results obtained in these three studies for the formation of benzene and toluene and the routes to their formation. The use of methane as the background gas makes these flames more representative of the combustion of a real fuel compared to a flame containing an unsaturated soot precursor as the only reactant.

# EXPERIMENTAL PROCEDURE AND MAIN RESULTS

The experiments were performed using a horizontal laminar premixed flat flame burner housed in a water-cooled vacuum chamber evacuated by two primary pumps and maintained at 6.7 kPa (50 Torr) by a regulation valve [12]. This chamber was equipped with a silica microprobe for sampling and a thermocouple for temperature measurements. The position of the burner could be vertically adjusted, while the housing and its equipments were kept fixed.

Gas flow rates were controlled by mass flow regulators. The flow rate of liquid cyclopentene was controlled using a liquid mass flow controller, mixed with the carrier gas and then evaporated by passing through a single pass heat exchanger [14]. Temperature profiles were obtained using a PtRh (6%)-PtRh (30%) type B thermocouple (diameter 100 µm) coated with an inert layer to prevent catalytic effects. Radiative heat losses were corrected using the electric compensation method. Figure 1 presents the experimental temperature profiles for methane flames doped with allene, 1,3-butadiene and cyclopentene. The profiles measured with propyne is very close to that with allene. In the studied flames, the temperature ranged from 600 K near the burner up to 2150 K.

**FIGURE 1**

Gas samples were directly obtained by connecting through a heated line (temperature between 393 and 423 K) the quartz probe to a volume which was previously evacuated. The pressure drop between the flame at a pressure of 6.7 kPa (50 Torr) and the inlet of the probe at a pressure always below 1.3 kPa (10 Torr) ensured that reactions were frozen. Stable species profiles were determined by gas chromatography. Chromatographs at atmospheric pressure with a Carbosphere packed column and helium or argon as carrier gas were used to analyse $O_2$, $H_2$, CO and $CO_2$ by thermal conductivity detection and $CH_4$, $C_2H_2$, $C_2H_4$, $C_2H_6$ by flame ionisation detection (FID). Heavier hydrocarbons from $C_3$ to $C_7$ were analysed on a Haysep packed column by FID and nitrogen as carrier gas. The identification was performed using GC/MS and by comparison of retention times.

The experimental conditions used for the four flames are summarized in Table 1; $\phi$ was equal to 1.55 in the flames seeded with allene and propyne and to 1.79 in the two others. Due to problems of stability and in order to get a wider reaction zone above the burner, a slightly higher dilution yield has been used in the flame of cyclopentene compared to the 3 other flames. This higher dilution explains the lower temperature profile obtained with cyclopentene compared to 1,3-butadiene, as shown in fig. 1.

In the pure methane flame (containing 20.9% (molar) of methane and 33.4 % of oxygen corresponding to an equivalent ratio of 1.25 [12]) and in all the doped ones, quantified products included carbon monoxide and dioxide, hydrogen, ethane, ethylene, acetylene, propyne (p-$C_3H_4$), allene (a-$C_3H_4$), propene ($C_3H_6$) and propane ($C_3H_8$). Several $C_{3+}$ products which cannot be detected in the pure methane flame have also been analysed:

- **in the case of C₃ additives**, butadienes, 1-butene, iso-butene, 1-butyne, vinylacetylene and benzene (toluene could be quantified for flames of richer mixtures, but was not detected in the flames studied here) [12],

- **in the case of 1,3-butadiene**, 1,2-butadiene, butynes, vinylacetylene, diacetylene, 1,3-pentadiene, 2-methyl-1,3-butadiene (isoprene), 1-pentene, 3-methyl-1-butene, benzene and toluene [13],

- **in the case of cyclopentene**, butadienes, vinylacetylene, diacetylene, cyclopentadiene, 1,3-pentadiene, benzene and toluene [14].

In the four flames, the carbon balance calculated from the initial reactant mole fractions, the measured product mole fraction and the simulated argon mole fraction is better than $\pm$ 10% whatever the position in the flame.

## DESCRIPTION OF THE DETAILED KINETIC MODEL

The mechanism proposed to model the oxidation of allene, propyne, 1-3-butadiene and cyclopentene (see supplementary file in the website of the journal) includes a mechanism built to model the oxidation of $C_3$-$C_4$ unsaturated hydrocarbons [12-13], our previous mechanisms for the oxidation of benzene [15], toluene [16] and cyclopentene [14]. Thermochemical data were estimated using software THERGAS [17], which is based on the additivity methods proposed by Benson [18].

*Reaction base for the oxidation of $C_3$-$C_4$ unsaturated hydrocarbons [12-13]*

This $C_3$-$C_4$ reaction base [12-13] was built from a review of the literature. The $C_3$-$C_4$ reaction base includes reactions involving $C_3H_2$, $C_3H_3$, $C_3H_4$ (allene and propyne), $C_3H_5$, $C_3H_6$, $C_4H_2$, $C_4H_3$, $C_4H_4$, $C_4H_5$, $C_4H_6$ (1,3-butadiene, 1,2-butadiene, methyl-cyclopropene, 1-butyne and 2-butyne), $C_4H_7$ (6 isomers), as well as some reactions for linear and branched $C_5$ compounds and the formation of benzene. In this reaction base, pressure-dependent rate constants follow the formalism proposed by Troe [19] and efficiency coefficients have been included.

*Mechanisms for the oxidation of benzene and toluene*

Our mechanism for the oxidation of benzene contains 135 reactions and includes the reactions of benzene, cyclohexadienyl, phenyl, phenylperoxy, phenoxy, hydroxyphenoxy, cyclopentadienyl, cyclopentadienoxy and hydroxycyclopentadienyl free radicals, as well as the reactions of ortho-benzoquinone, phenol, cyclopentadiene, cyclopentadienone and vinylketene which are the primary products yielded [15].

The mechanism for the oxidation of toluene contains 193 reactions and includes the reactions of toluene, benzyl, tolyl, peroxybenzyl (methylphenyl), alcoxybenzyl and cresoxy free radicals, as well as the reactions of benzaldehyde, benzyl hydroperoxyde, cresol, benzylalcohol, ethylbenzene, styrene and bibenzyl [16]. To avoid an overprediction of toluene in the $C_3$ and $C_4$ flames, the rate constant of the decomposition of this molecule to give phenyl and methyl radicals, which is not well known, has been taken equal to the value proposed by Colket and Seery [20].

*Mechanism proposed for the oxidation of cyclopentene [14]*

We have considered the unimolecular reactions of cyclopentene, the additions of H-atoms and OH radicals to the double bonds and the H-abstractions by oxygen molecules and small radicals. Unimolecular reactions include dehydrogenation to give cyclopentadiene, decompositions by breaking of a C-H bond and isomerization to give 1,2-pentadiene. The reactions of cyclopentenyl radicals involve isomerizations, decompositions by breaking a C-C bond to form linear $C_5$ radicals including two double bonds or a triple bond and the formation of cyclopentadiene by breaking of a C-H bond or by oxidation with oxygen molecules and terminations steps. Termination steps are written only for resonance stabilized cyclopentenyl radicals: disproportionations with H-atoms and OH radicals yield cyclopentadiene, combinations with $HO_2$ radicals lead to ethylene and $CH_2CHCO$ and OH radicals (after decomposition of the obtained hydroperoxide) and combinations with $CH_3$ radicals form methylcyclopentene. The decomposition by breaking a C-H bond of cyclopentyl radicals lead to the formation of 1-penten-5-yl radicals. The isomerizations (for the radical stabilized isomer) and the decompositions by breaking of a C-C bond of the linear $C_5$ radicals were also written, while those by breaking of a C-H bond are not considered. The reactions of cyclopentadiene are part of the mechanism for the oxidation of benzene, but reactions for the consumption of methylcyclopentene and methylcyclopentadiene, obtained by recombination of cyclopentadienyl and methyl radicals had to be added.

*Ways of formation of aromatic compounds [12-14]*

In order to investigate the relative importance of the different channels, the formation of aromatic compounds was considered through the $C_3$, the $C_4$ and the $C_5$ pathways:

- **For the $C_3$ pathway**, we have used a value of $1x10^{12}$ cm$^3$.mol$^{-1}$s$^{-1}$ for the recombination of two propargyl radicals to give phenyl radicals and H-atoms, which is in good agreement with that recently proposed by Miller and Klippenstein [21] and Rasmussen et al. [22]. Note that because of the low pressure of our study, we have considered the formation of phenyl radicals and H-atoms to be the dominant channel from the recombination of propargyl radicals. Due to addition of the toluene mechanism including the ipso-addition of toluene with H-atoms leading to benzene and methyl radicals and to avoid an overprediction of benzene in the case of the flame doped with allene, we have removed from our previous mechanism [12] the addition of propargyl radicals to allene leading to benzene and H-atoms and the recombination between allene and propargyl radicals to produce phenyl radicals and two H-atoms. These reactions were not considered by Rasmussen et al. [22].

- **For the $C_4$ pathway**, all the reactions between $C_2$ species and n-$C_4H_3$, n-$C_4H_5$ radicals or 1,3-butadiene molecules and leading to aromatic and linear $C_6$ species have been considered as proposed by Westmoreland et al. [2]. We have written the main reactions of 1,4-cyclohexadiene: its dehydrogenation to give benzene with a rate constant proposed by Ellis and Freys [23] and the H-abstractions by H-atoms and OH radicals to give $C_6H_7$ radicals with rate constants proposed by Dayma et al. [24].

- **For the C$_5$ pathway**, we have considered the isomerization between both radicals deriving from methylcyclopentadiene by H-abstractions (cyclopentadienylmethylene and methylcyclopentadienyl) and the formation of cyclohexadienyl radicals from cyclopentadienyl methylene radicals as proposed by Lifshitz et al. [10]. As it was not considered by Lifshitz et al. [10], we have not written the formation of cyclohexadienyl radicals from methyl cyclopentadienyl radicals as proposed by Marinov et al. [4], but the recombination with H-atoms of these resonance stabilized radicals was also added. In order to reproduce the important formation of toluene experimentally observed in the flame doped with cyclopentene, we have considered the formation of benzyl radicals from the addition of acetylene to cyclopentadienyl radicals with a rate constant about 3 times larger than that usually considered for such an addition [25]. While the decomposition of benzyl radicals to form acetylene and cyclopentadienyl radicals, which was considered in our mechanism for the oxidation of toluene, has been studied by several authors [26-27] and has been shown to occur through a several-step mechanism, the reverse addition has never been directly investigated.

**COMPARISON OF THE FORMATION OF BENZENE AND TOLUENE IN THE FOUR FLAMES**

All the simulations have been performed using PREMIX of CHEMKIN II [28] with estimated transport coefficients, the same kinetic mechanism and using the experimental temperature profile as an input. To compensate the perturbations induced by the quartz probe and the thermocouple, the temperature profile used in calculations is an average between the experimental profiles measured with and without the quartz probe, with a slight shift away from the burner surface, as shown in figure 1.

Comparisons between experimental results and simulations are displayed for benzene in fig. 2 and for toluene in fig. 3: the unique model reproduces correctly the experimental formation of these two aromatic compounds in the four flames. Compared to our previous simulation with less complete mechanisms [12-13], the simulated maximum mole fractions of benzene have been increased by a factor 1.08, 1.10 and 1.48, for the flame of allene, propyne and 1,3-butadiene, respectively. A detectable concentration of toluene is predicted, but not observed experimentally for the flames containing $C_3$ additives and the production of toluene is underestimated by a factor 3 for the flame doped with cyclopentene.

**FIGURES 2 AND 3**

The experimental results show important changes in the formation of aromatic compounds depending on the additives. The production of aromatic compounds occurs the closest to the burner in the flames doped by $C_3$ compounds and the furthest in the flame seeded by cyclopentene. As the concentration of obtained aromatic species depends strongly on equivalence ratio, Table 1 presents the maximum mole fractions of benzene and toluene simulated for all the flames under the same conditions as in the flame of cyclopentene. Note however that the fact to use the same temperature profile for all the flames as for the flame of cyclopentene involves a slight overestimation of temperature. The flame showing the largest formation of benzene is the one doped with cyclopentene. Under the same conditions, the simulated maximum of the peak of benzene is 3.8 times smaller in the flame doped with 1,3-butadiene, 5.5 times in the flame doped by allene and 6.6 times in that doped by propyne. The difference between the flames doped by allene and propyne is connected to differences in the rates of formation and consumption of propargyl radicals between both flames [12]. The flame doped with 1,3-butadiene exhibits the lowest production of toluene. The simulated peak of toluene would be around 4 times larger in the flames doped with $C_3$ compounds and almost 25 times in that doped with cyclopentene.

**DISCUSSION**

Figure 4 presents a reaction rate analysis for the production and consumption of aromatic compounds for the flames doped with allene (a similar picture is obtained for propyne), 1,3-butadiene and cyclopentene. Figure 5 presents a sensitivity analysis for the reactions for which a variation of the rate constant has an important effect on the formation of benzene and toluene.

These figures show well that the ways of formation of aromatic compounds are very different depending on the additive.

<div align="center">**FIGURES 4 AND 5**</div>

In the allene flame, phenyl radicals are mainly obtained by self-combination of propargyl radicals. Combinations of phenyl radicals with H-atoms and methyl radicals lead to benzene and toluene, respectively. As they derive directly from propargyl and phenyl radicals, benzene and toluene are primary products. That explains why their formation occurs early in the flame, as propargyl radicals are directly obtained from allene and propyne by H-abstractions. The production of benzene is relatively weak because the self-combination of propargyl radicals, at the low pressure studied here, leads mainly to phenyl radicals which react rapidly with oxygen molecules. As shown in figure 4, the rate of the reaction of phenyl radicals with oxygen molecules to give phenoxy radicals is equal to six times that of the formation of benzene. It is worth noting that our simulation using a value of the rate constant for the self-combination of propargyl radicals close to that given in the literature leads to a good agreement for the formation of benzene. This rate constant has the most important sensitivity coefficient for the formation of benzene. Many similarities exist between allene and propyne flames in both reactivity and product formation. Nevertheless, the pool of small radicals is slightly larger in the propyne flame, which leads to a faster consumption of propargyl radicals and a lower formation of benzene than in the flame seeded by allene [12]. In this flame, the $C_5$ pathway accounts for 5% of the formation of benzene, since the decomposition of phenoxy radicals leads to cyclopentadienyl radicals. The contribution of the $C_4$ pathway is negligible. The reactions which have the largest influence on the formation of toluene are the combination of propargyl radicals and that of methyl and phenyl radicals (see fig. 5). This last reaction has also a promoting effect on the formation of benzene as 10% of this aromatic compound is obtained through the ipso-addition of H-atoms to toluene with elimination of methyl radicals.

Of particular interest in the flames doped by 1,3-butadiene and cyclopentene is that the production of benzene is mainly due to other reactions rather than to the $C_3$ pathway as it is often the case in the flames described in the literature [22]. Note however that the relative importance of the $C_4$ and $C_5$ pathway compared to the $C_3$ one considerably depends of the products assumed for the recombination of two propargyl radicals.

In the 1,3-butadiene flame, benzene is mainly formed from the addition/cyclization of vinyl radicals to produce cyclohexadiene which reacts either by dehydrogenation or by methatheses with H-atoms and OH radicals, followed by the decomposition of the obtained cyclic $C_6H_7$ radicals. While the rate constant having the most important sensitivity coefficient for the formation of benzene in the flame including the $C_4$ additive is that of the reaction between 1,3-butadiene and vinyl radicals, figure 4 points out also an important contribution of the $C_5$ pathway to the formation of cyclohexadienyl radicals (about 40%). This is explained by a production of cyclopentadiene deriving, via methylcyclopentenyl radicals, from a reaction between acetylene and the resonance stabilized 1-buten-3-yl radicals which are obtained by H-addition to 1,3-butadiene. These reactions which are included in the present mechanism explain the important increase of the formation of benzene compared to our previous mechanism [13]. Only 1% of the formation of benzene is due to the $C_3$ pathway, since allene and propyne are minor products of the oxidation of 1,3-butadiene.

In the cyclopentene flame, the major source of benzene is the isomerization of cyclopentadienyl methylene radicals which are obtained either directly by H-abstractions by H-atoms or OH radicals from methylcyclopentadiene or by isomerisation from the resonance stabilized methyl cyclopentadienyl radicals. In this flame, the most important reaction for the formation of methylcyclopentadienyl radicals is the unimolecular decomposition of methylcyclopentadiene by breaking of a C-H bond. This decomposition has the most important sensitivity coefficient for the formation of benzene. The H-abstractions by H-atoms and OH radicals are only minor channels. Methylcyclopentadiene is obtained from the recombination of methyl and cyclopentadienyl radicals, these last radicals deriving from cyclopentadiene are a major product of the oxidation of cyclopentene. While the contribution of the $C_4$ pathway is negligible, that of the $C_3$ pathways is 7% of the total production of benzene. Propargyl radicals are produced by beta-scission from 1-pentyn-5yl radicals, which are obtained by ring opening from vinylic cyclopentenyl radicals.

In the 1,3-butadiene and cyclopentene flames, the most important way to give toluene is the addition of resonance stabilized cyclopentadienyl radicals to acetylene. Consequently, the reactions which have the largest influence on the formation of toluene are the addition of cyclopentadienyl radicals to acetylene giving benzyl radicals and the combination of these last radicals with H-atoms. In the 1,3-butadiene flame, as the flow of formation of toluene is around 10 times lower than that of benzene (see figure 4), the combination of methyl and phenyl radicals is also of some importance. Whatever the type of flames, cyclopentadienyl radicals are rapidly obtained from the decomposition of phenoxy radicals, which are produced by the reaction of phenyl radicals with oxygen molecules. Due to their resonance stabilization, they are present with an important concentration in every system containing phenyl radicals and benzene. The formation of toluene by addition of cyclopentadienyl radicals to acetylene could then be of importance in many combustion systems rich in acetylene.

Other aromatic molecules, which are predicted by simulation but could not be detected experimentally due to the chromatographic column used, are phenol, benzaldehyde and ethylbenzene. Phenol is obtained by combination of H-atoms and phenoxy radicals. Benzaldehyde and ethylbenzene derived from benzyl radicals by reaction with O-atoms and methyl radicals, respectively, and their degradation reactions produce phenyl radicals and benzene, respectively. In the cyclopentene flame, 5% of benzene is formed via ethylbenzene.

**CONCLUSION**

This paper compares the formation of benzene and toluene in a rich ($\Phi=1.8$) methane flame consecutively seeded with allene, propyne, 1,3-butadiene, and cyclopentene and shows that the amount of aromatic compounds formed are very different depending on the additive. This effect is due to differences in the pathways of formation of these aromatic species depending on the dopant.

The changes in the predicted concentrations of benzene and toluene encountered when using evolving versions of the mechanism from our first paper [12] up until the one presented here clearly demonstrates the need to use the most detailed models so that no important reaction is omitted and the degree of coupling between the reactions involved in the different pathways leading to aromatic compounds. The reaction rates and sensitivity analyses presented in this paper allow these reaction links to be better identified.

**Table 1**: Experimental flame characteristics: initial flow velocities at 333 K (in cm/s), equivalence ratios ($\phi$), initial reactant mole fractions (X(Ar)$_0$, X(O$_2$)$_0$, X(CH$_4$)$_0$, X(additive)$_0$) and C/H and C/O ratios, and maximum mole fractions of benzene and toluene (X(benzene)$_{max}$ and X(toluene)$_{max}$) simulated under the same conditions ($\phi$, dilution yield, initial additive mole fraction and temperature profile) as in the flame seeded by cyclopentene.

| Additive | allene | propyne | 1,3-butadiene | cyclopentene |
|---|---|---|---|---|
| Initial flow velocity | 36 | 36 | 36 | 36 |
| $\phi$ | 1.55 | 1.55 | 1.79 | 1.79 |
| X(Ar)$_0$ | 0.432 | 0.432 | 0.423 | 0.556 |
| X(O$_2$)$_0$ | 0.334 | 0.334 | 0.331 | 0.267 |
| X(CH$_4$)$_0$ | 0.209 | 0.209 | 0.207 | 0.153 |
| X(additive)$_0$ | 0.025 | 0.025 | 0.033 | 0.024 |
| C/H | 0.303 | 0.303 | 0.330 | 0.340 |
| C/O | 0.425 | 0.425 | 0.512 | 0.511 |
| X(benzene)$_{max}$ | $1.56 \times 10^{-5}$ | $1.21 \times 10^{-5}$ | $3.33 \times 10^{-5}$ | $1.27 \times 10^{-4}$ |
| X(toluene)$_{max}$ | $8.78 \times 10^{-5}$ | $7.51 \times 10^{-6}$ | $2.00 \times 10^{-6}$ | $5.00 \times 10^{-5}$ |

**FIGURE CAPTIONS**

**Figure 1**: Experimental (symbols) temperature profiles measured without the probe and temperature profiles (lines) used for simulations for the allene, 1,3-butadiene and cyclopentene flames.

**Figure 2**: Experimental (symbols) and simulated (lines) profiles of benzene in a methane flame doped with (a) allene (black symbols and full line) and propyne (white symbols and broken line), (b) 1,3-butadiene and (c) cyclopentene.

**Figure 3**: Experimental (symbols) and simulated (lines) profiles of toluene in a methane flame doped with (a) allene (full line) and propyne (broken line), (b) 1,3-butadiene and (c) cyclopentene.

**Figure 4**: Reaction rates analysis for the formation and consumption of benzene and toluene for a position in the flame corresponding to the peak of benzene profile. The numbers (normal typing for allene (T = 1218 K), italic for 1,3-butadiene (T = 1393 K) and bold for cyclopentene (T = 1354 K)) are the rates normalized by the rate of formation of benzene in each flame.

**Figure 5**: Variation of the normalized (compared to the value obtained with the initial mechanism) maximum mole fractions of benzene and toluene when multiplying by a factor of 2 the rate constant of the considered reaction (initial rate constant are given in mol, $cm^3$, s units). Only reactions for which a variation above 1.3 is obtained in at least one of the four flames are displayed.

**LIST OF SUPPLEMENTAL MATERIAL :**

1 supplemental file: mechanism-gueniche et al.-2007

**CAPTION FOR SUPPLEMENTAL MATERIAL :**

Mechanism (in CHEMKIN format) for the oxidation of allene, propyne, 1,3-butadiene and cyclopentene in low pressure flames.

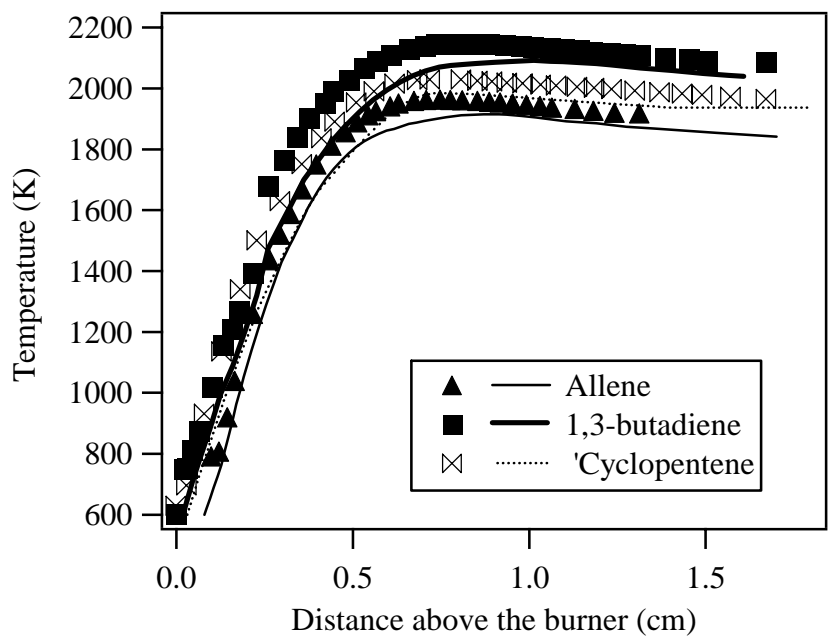

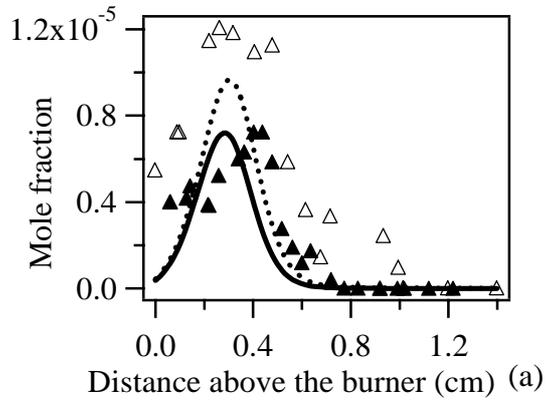

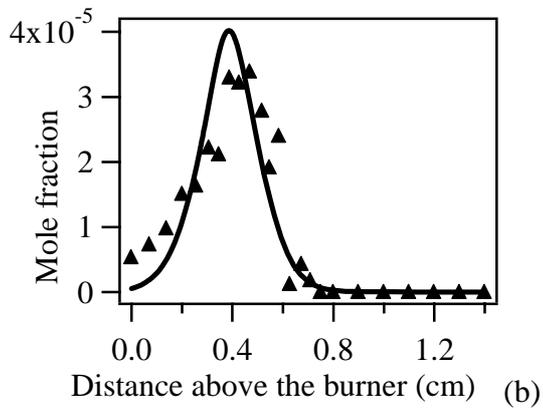

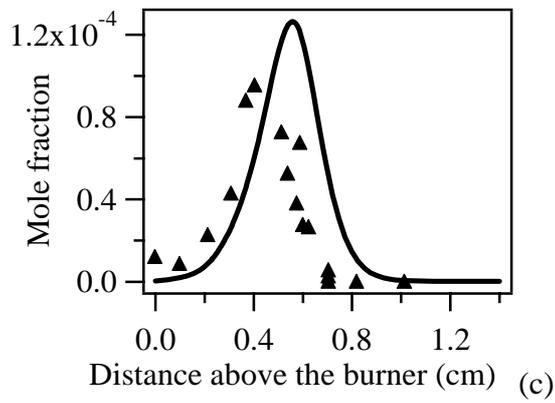

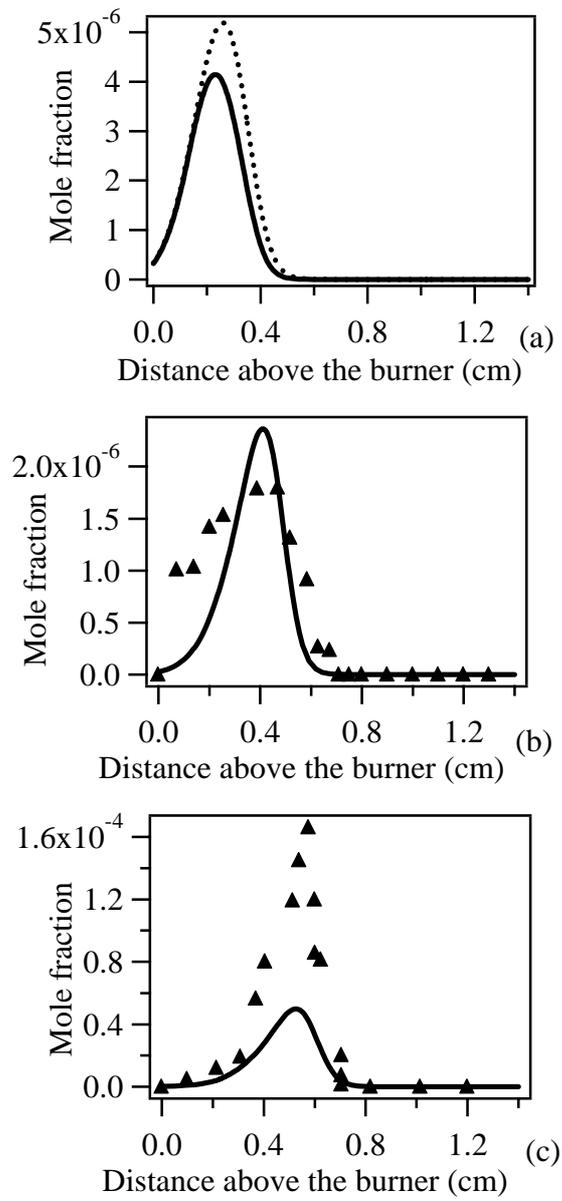

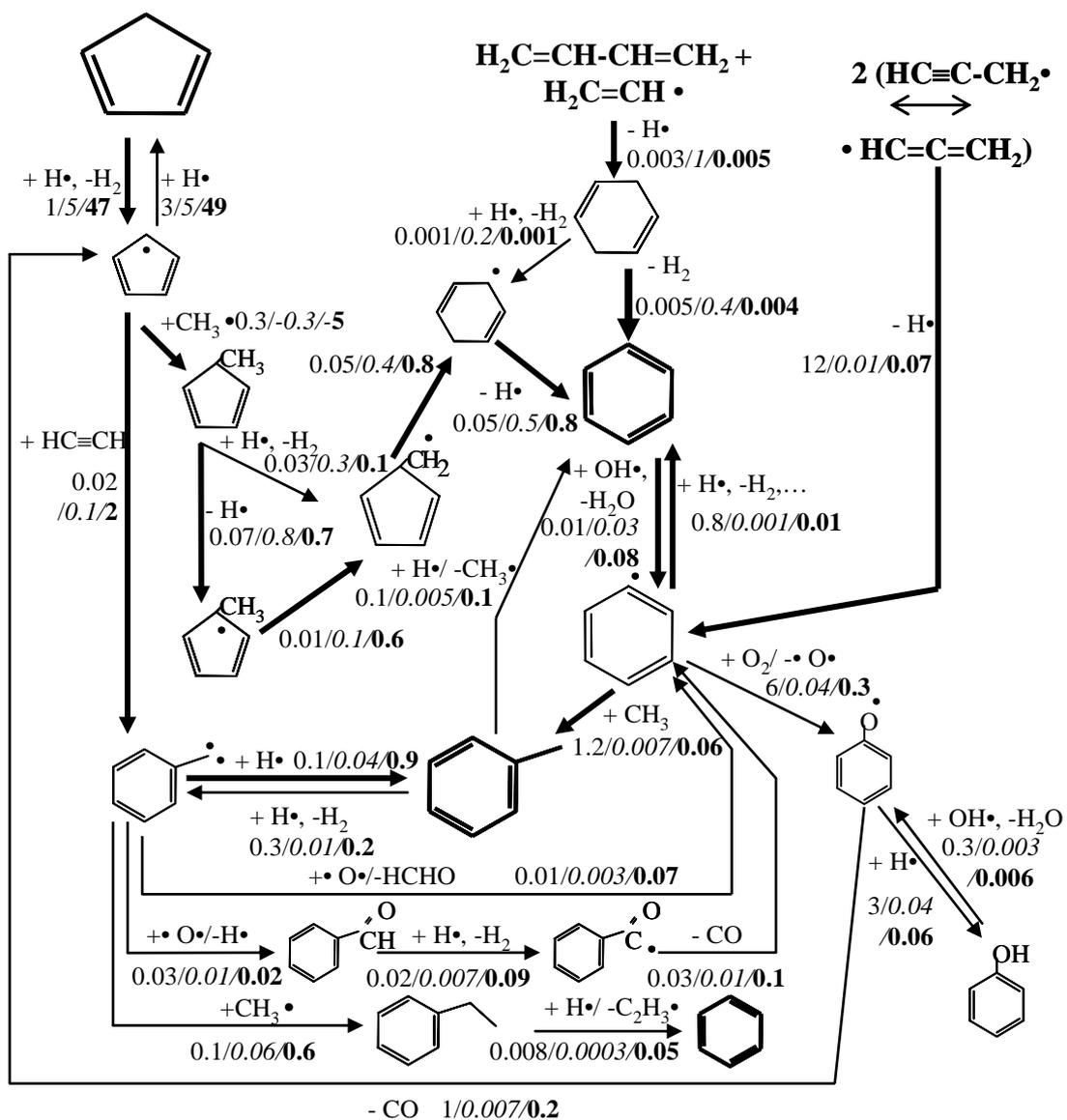

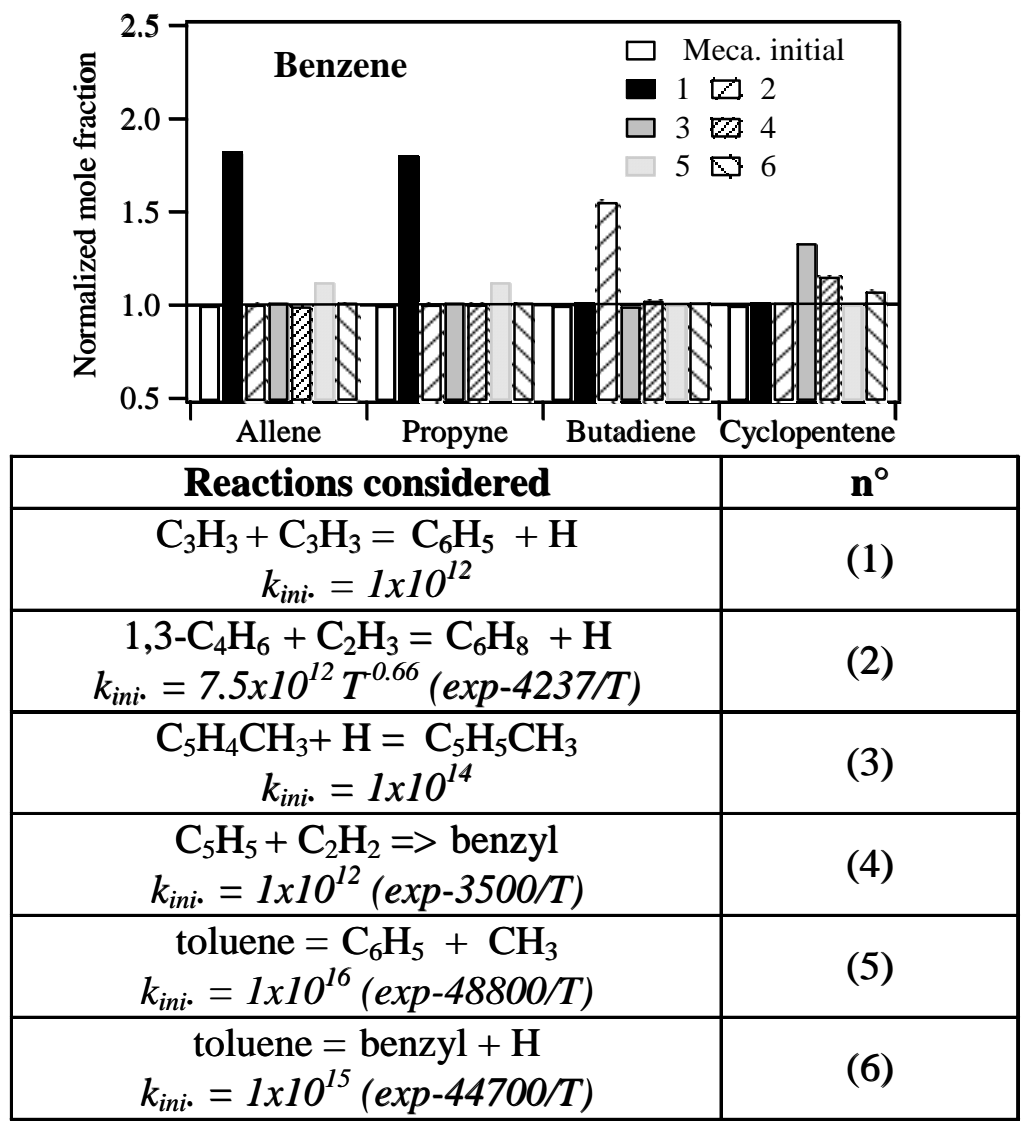

| Reactions considered | n° |
|---|---|
| $C_3H_3 + C_3H_3 = C_6H_5 + H$<br>$k_{ini.} = 1x10^{12}$ | (1) |
| $1,3\text{-}C_4H_6 + C_2H_3 = C_6H_8 + H$<br>$k_{ini.} = 7.5x10^{12} T^{-0.66} (exp\text{-}4237/T)$ | (2) |
| $C_5H_4CH_3 + H = C_5H_5CH_3$<br>$k_{ini.} = 1x10^{14}$ | (3) |
| $C_5H_5 + C_2H_2 => benzyl$<br>$k_{ini.} = 1x10^{12} (exp\text{-}3500/T)$ | (4) |
| $toluene = C_6H_5 + CH_3$<br>$k_{ini.} = 1x10^{16} (exp\text{-}48800/T)$ | (5) |
| $toluene = benzyl + H$<br>$k_{ini.} = 1x10^{15} (exp\text{-}44700/T)$ | (6) |

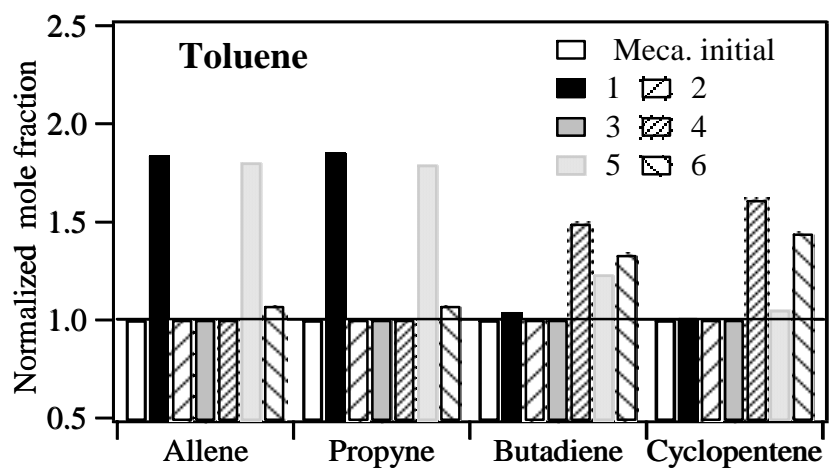